\documentstyle[epsfig]{mn}
\begin{document}
\def\ltsima{$\; \buildrel < \over \sim \;$}
\def\simlt{\lower.5ex\hbox{\ltsima}}
\def\gtsima{$\; \buildrel > \over \sim \;$}
\def\simgt{\lower.5ex\hbox{\gtsima}}
\def\approxgt{\mathrel{\hbox{\rlap{\lower.55ex \hbox {$\sim$}}
        \kern-.3em \raise.4ex \hbox{$>$}}}}
\def\approxlt{\mathrel{\hbox{\rlap{\lower.55ex \hbox {$\sim$}}
        \kern-.3em \raise.4ex \hbox{$<$}}}}

\title[]{First hard X-ray detection  and broad band X-ray study of  the unidentified transient AX J1949.8+2534} 

\author[Sguera et al.]
{V. Sguera$^{1}$, L. Sidoli$^2$, A. Paizis$^2$, N. Masetti$^{1}$, A. J. Bird$^3$, A. Bazzano$^4$\\ 
$^1$ INAF, Istituto di Astrofisica Spaziale e Fisica Cosmica, Via Gobetti 101, I-40129 Bologna, Italy \\
$^2$ INAF, Istituto di Astrofisica Spaziale e Fisica Cosmica, Via E. Bassini  15, I-20133 Milano, Italy \\
$^3$ School of Physics and Astronomy, University of Southampton, University Road, Southampton, SO17 1BJ, UK \\
$^4$ INAF, Istituto di Astrofisica e Planetologia Spaziali, Via Fosso del Cavaliere 100, I-00133 Roma, Italy \\
}

\date{Accepted 2017 May 05. In original form 2017 March 01}

\maketitle

\begin{abstract}
We present the results from  {\itshape INTEGRAL} and {\itshape Swift/XRT} observations of the hitherto  poorly studied unidentified X-ray transient
AX J1949.8+2534,  and on  archival multiwavelength  observations of field objects. Bright hard X-ray outbursts have been discovered  above 20 keV for the first time, the measured duty cycle  and  dynamic range are of the order of 
$\sim$ 4\% and $\ge$ 630, respectively. The source was also detected during a low soft X-ray state ($\sim$ 2$\times$10$^{-12}$ erg cm$^{-2}$ s$^{-1}$) thanks to a {\itshape Swift/XRT}  followup, which  allowed for the first time to perform a soft X-ray spectral analysis as well as significantly improve the source positional uncertainty from arcminute to arcsecond size. From archival near-infrared data, we pinpointed two bright objects as most likely counterparts whose   photometric properties are compatible with an early type  spectral nature. This strongly supports  a High Mass X-ray Binary (HMXB) scenario for AX~J1949.8+2534, specifically a Supergiant Fast X-ray Transient (more likely) or alternatively a Be HMXB.
\end{abstract}


\vspace{1.0cm}

\section{Introduction}
AX J1949.8+2534 is a hitherto poorly studied unidentified X-ray source. It was discovered during the 
{\itshape ASCA} Galactic Plane Survey (Sugizaki et al. 2001) whose observations were carried out from March 1996  to  April 1999, covering the inner Galactic disk at $\left|l\right|$$\le$45$^\circ$
and  $\left|b\right|$$\le$0$^\circ$.4. The source was detected with a significance of 3.6$\sigma$ and 21.6$\sigma$ in the energy bands 
0.7--2 keV and 2--10 keV, respectively.  No  spectral information is available, the reported 2--10 keV
{\itshape ASCA} count rate  (Sugizaki et al. 2001) converts into  an absorbed X-ray flux  of $\sim$ 8$\times$10$^{-12}$ erg cm$^{-2}$ s$^{-1}$ ($\sim$ 6$\times$10$^{-12}$ erg cm$^{-2}$ s$^{-1}$) if we assume a power law spectral shape with $\Gamma$=1 ($\Gamma$=2). The uncertainty on the source  position is $\sim$ 1$'$ as is typical for {\itshape ASCA} X-ray sources, this  prohibits the search for counterparts at lower energies (i.e. optical and infrared) which is essential to firmly identify the nature of the source. 

For many years,  the {\itshape ASCA} detection represented  the only information available  in the literature.  Interestingly, Sguera et al. (2015) have recently reported the first hard X-ray detection of AX J1949.8+2534 above 20 keV. Their communication contained only very short information about the hard   X-ray activity detected by {\itshape INTEGRAL},  as obtained from analysis  of near real time data  pertaining to public observations  of  the Cygnus region.  

Here we present  the results of a  more detailed spectral and temporal analysis of the consolidated {\itshape INTEGRAL} data  pertaining to the 
outburst  reported by Sguera et al. (2015), together with the investigation of additional  archival  {\itshape INTEGRAL}
data, with the aim of finding  further hard X-ray activity from the source. We also report a Target of Opportunity (ToO) observation made with  the {\itshape Swift} satellite in order to refine the error circle to   arcsecond size as well as to characterize for the first time the spectral shape in the 
soft X-ray band.

\section{{\itshape INTEGRAL}}

\subsection{Data analysis}

\begin{table}
\caption {Log of IBIS/ISGRI observations used for our study on AX~J1949.8+2534. Orbits in boldface  contain the IBIS/ISGRI detections reported in section 2.2.   $\dagger$  Detection originally reported by Sguera et al. (2015). } 
\label{tab:main_outbursts} 
\begin{tabular}{lccc}
\hline
\hline   
Telescope  &       Date                        &           Observation        &            Exposure       \\
Orbit           &                              &                    Target             &              (ks)        \\
\hline    
1600       &  18-20 Oct  2015          &          Cyg X-1      &      $\sim$  22  \\   
1601       &  22-23 Oct  2015       &              Cyg X-1   &      $\sim$  8  \\  
1602      &  23-25 Oct  2015       &  GPS+Cyg X-1   &      $\sim$  14  \\  
1603       &  26-28 Oct  2015      &  GPS+Cyg X-1  &      $\sim$  7  \\  
{\bfseries  1605$\dagger$  }      &  31 Oct - 02 Nov  2015      &  GPS+Cyg X-1&      $\sim$  16  \\ 
1606       &  03  Nov  2015        &  GPS                   &      $\sim$  3.5  \\ 
1607      &  05-07  Nov  2015         &  GPS+Cyg X-1    &      $\sim$  22  \\ 
1609       &  12-13  Nov  2015            &   Cyg X-1 &      $\sim$  6  \\ 
1610       &  14-15  Nov  2015       &   Cyg X-1&      $\sim$  12  \\ 
1611        &  18  Nov  2015             &  GPS    &      $\sim$  3.5  \\ 
1613       &  23  Nov  2015                 &  GPS &      $\sim$  4  \\ 
1614       &  24-26  Nov  2015            &  GPS+Cyg X-1    &      $\sim$  20  \\ 
1616        &  01  Dec  2015                &  GPS+Cyg X-1    &      $\sim$  7  \\ 
1618         &  05  Dec  2015     &  GPS   &    $\sim$  4  \\ 
1619          &  09  Dec  2015       &  GPS   &      $\sim$  4  \\ 
1621          &  13-15  Dec  2015   & Cyg X-1&      $\sim$  15  \\ 
1624           &  21-23  Dec  2015        &  GPS+Cyg X-1    &      $\sim$  18  \\
1626           &  26-28  Dec  2015           &  ToO V404 Cyg   &      $\sim$  17  \\
1627           &  29-31  Dec  2015          &  ToO V404  Cyg  &      $\sim$  16  \\
{\bfseries 1628}            &  31  Dec - 02 Jan 2016     &  ToO V404  Cyg  &      $\sim$  18  \\
{\bfseries 1629}           &  03-05  Jan  2016            &  ToO V404    Cyg &      $\sim$  19  \\
\hline
                                       &               &      &      $\sim$  255  \\  
\hline
\hline  
\end{tabular}
\end{table} 

For our study, we used data collected with the ISGRI detector (Lebrun et al. 2003) which is  the lower energy 
layer of the IBIS coded mask telescope (Ubertini et al. 2003) onboard {\itshape INTEGRAL} (Winkler et al. 2003). 
The  reduction  and  analysis  of  the data  have been  performed  by  using  the  Offline  Scientific Analysis  (OSA)  version 10.1. For IBIS/ISGRI spectral analysis we used the standard 13 energy channel response matrix available at the {\itshape INTEGRAL} Science Data Centre (ISDC).   {\itshape INTEGRAL} observations  during  each orbit ($''$revolution$''$, lasting $\sim$ 2.6 days or 170 ks) are  divided  into short pointings (Science Window, ScW) having  a typical duration of $\sim$ 2,000 s. 

Our total data set consists of all public observations which covered  the Cygnus region (i.e. Galactic Plane Scans GPS, ToO observations of V404 Cyg and targeted observations of Cyg X--1) immediately before and after the hard X-ray detection of AX~J1949.8+2534  reported by Sguera et al. (2015). The corresponding data set  amounts to a total exposure of $\sim$ 255 ks (see Table 1 for details). 
 
We performed an analysis of the full data set on two different timescales, i.e. at ScW level as well as at revolution level,  in order to  search for 
newly discovered  X-ray activity from the source detected with a significance equal or greater  than at least 5$\sigma$ and 7$\sigma$, respectively. Such detection thresholds  are essential to avoid false detections/excesses caused by background noise (e.g. Bird et al. 2016). 
The search was initially performed in the energy band 22--60 keV;  this choice takes into account the evolution 
of the IBIS/ISGRI energy threshold  that occurred from revolution  number  $\sim$ 900 on. When a significant detection  was found,  we have also checked the detection at higher energies (i.e. 60--100 keV) or in other  different  ranges (i.e. 22--30, 30--60 and 22--40 keV). 
We note that the  sensitivity limit for a persistent source detected at 5$\sigma$ level (22--60 keV) in only one ScW of
about 2,000 s duration is  $\sim$ 18 mCrab (Krivonos et al. 2010). 
 
The X-ray monitor JEM--X   (Lund et al. 2003) makes observations simultaneously with
IBIS/ISGRI, although with a much smaller Field of View (FoV),  providing images in the softer energy band 3--35 keV. 
JEM--X data were analyzed when the source was  in its FoV in order  to search for X-ray activity  in 
both the energy bands 3--10 keV and 10--20 keV.  

Throughout the paper, the spectral analysis was performed using
XSPEC version 12.9.0 and, unless stated otherwise, errors are quoted at the 90 per cent confidence
level for one single parameter of interest.

\subsection{Results}

We report  on newly discovered  hard X-ray transient activity from  AX J1949.8+2534, the first  ever  above 20 keV.  Hard X-ray detections  with IBIS/ISGRI were obtained by analyzing data in revolutions number 1605, 1628 and 1629. Table 2 reports a summary of the outbursts main characteristics.

\begin{table}
\caption {List of  {\itshape INTEGRAL}  orbits during which significant source detections (i.e. $\ge$ 7$\sigma$)  were obtained in the energy band 22--60 keV. The table also lists the date of the beginning of the outburst,  significance detection over  the
entire activity, average flux and approximate duration.} 
\label{tab:main_outbursts} 
\begin{tabular}{ccccc}
\hline
\hline    
Orbit    &     Date                          &    Significance     &    Flux             &  Duration       \\
    (n.)                   &  (MJD)                        &   ($\sigma$)            &    (mCrab)          & (days)  \\
\hline    
1605                    &   57327.35   &  7.1            &     10.0$\pm$1.4                      & $\sim$ 1.5      \\  
1628                    &   57387.65                    &    7.1        &       10.6$\pm$1.5     &  $\sim$ 2   \\                    
1629                    &    57390.31              &     7.3                 &    9.5$\pm$1.3      &  $\sim$ 2 \\ 
\hline
\hline  
\end{tabular}
\end{table} 

\subsubsection{IBIS/ISGRI detection in revolution 1605}

Firstly, we note that AX J1949.8+2534 was not detected in any single revolution from n. 1600 to n. 1603, nor in their mosaic for a total on source exposure of $\sim$ 50 ks.  As a result, we inferred a 3$\sigma$ upper limit of $\sim$ 2 mCrab (22--60 keV).

Conversely, AX J1949.8+2534 was detected with a significance of  7.1$\sigma$ (22--60 keV) during revolution 1605 ($\sim$ 14 ks on-source exposure)  from 2015 Nov 01 08:28 (UTC) to 2015 Nov 02 15:48 (UTC). No detection was obtained in the energy band 60--100 keV. The measured average 22--60 keV flux is 10.0$\pm$1.4 mCrab (or $\sim$ 1.1$\times$10$^{-10}$ erg cm $^{-2}$ s$^{-1}$). The source was never significantly detected  at ScW level  (i.e. $\ge$ 5$\sigma$) at any point  of the observation, indicating that no major flaring  activity took place on short timescales (i.e. $\sim$ 2,000 s).  

The extracted IBIS/ISGRI spectrum  was  fitted by a power law with $\Gamma$=2.9$\pm$0.8  ($\chi^{2}_{\nu}$=1.7, 4 d.o.f.) 
or alternatively by a  thermal bremsstrahlung with   kT=24$^{+22}_{-9}$ keV ($\chi^{2}_{\nu}$=1.5, 4 d.o.f.). 
The best fit was achieved by using a black body model  ($\chi^{2}_{\nu}$=1.15, 4 d.o.f.)   with 
kT=7.9$^{+1.8}_{-1.9}$ keV. The average 18--60 keV (20--40 keV) flux is 
1.1$\times$10$^{-10}$ erg cm $^{-2}$ s$^{-1}$ (7.8$\times$10$^{-11}$ erg cm $^{-2}$ s$^{-1}$). 
Fig. 1  shows the black body data-to-model fit with the corresponding residuals.  

The source was also in the JEM--X FoV during this observation, however in the combined JEM--X1+JEM--X2 mosaic it was not  detected in both  bands 3--10 keV and 10--20 keV  (on-source exposure of $\sim$ 4.8 ks). The inferred 3$\sigma$ upper limit (3--10 keV) is of the order of $\sim$ 2 mCrab or 4.4$\times$10$^{-11}$ erg cm $^{-2}$ s$^{-1}$.

 AX~J1949.8+2534 was not detected in any single revolution after n. 1605 (from n. 1606 
 to n. 1627), nor in their mosaic for a total on source exposure of $\sim$ 150 ks.  We inferred a 3$\sigma$ upper limit of $\sim$ 1.2 mCrab (22--60 keV). This, combined with the other upper limit from revolutions n. 1600 to 1603,  allows us to confidently constrain the duration of the transient hard X-ray activity detected in revolution n. 1605 to no longer than $\sim$ 1.5 days.

\begin{figure}
\begin{center}
\includegraphics[height=8.cm,angle=270]{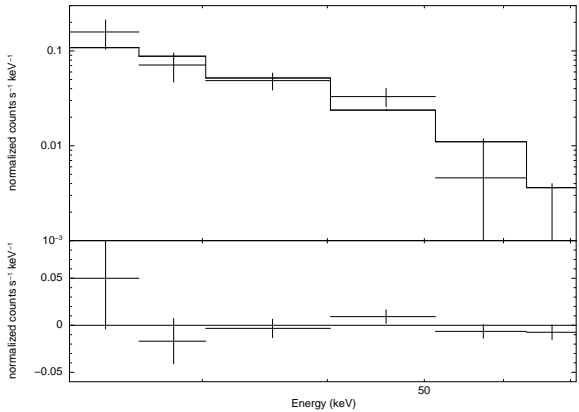}
\caption{IBIS/ISGRI spectrum of   AX~J1949.8+2534 (during revolution n. 1605) best fitted by a black body. The lower panel shows
the residuals from the fit. }
\label{fig6}
\end{center}
\end{figure}

\begin{figure}
\begin{center}
\includegraphics[height=8.5cm,angle=270]{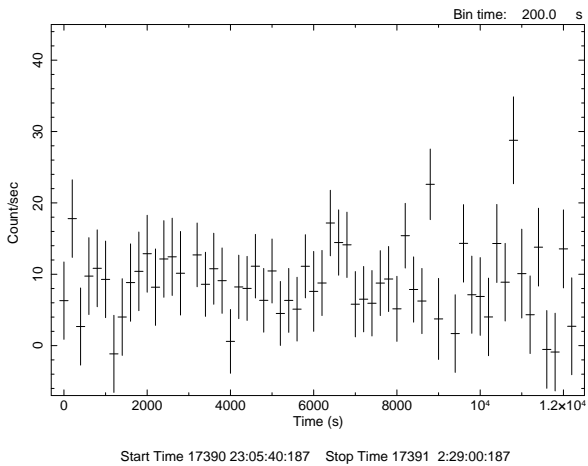}
\caption{IBIS/ISGRI light curve (22--60 keV, 200 s bin time) of  AX J1949.8+2534  extracted from the four consecutive ScWs n. 20 to 23 in revolution 1629.}
\label{fig6}
\end{center}
\end{figure} 

\begin{figure}
\begin{center}
\includegraphics[height=8.cm,angle=270]{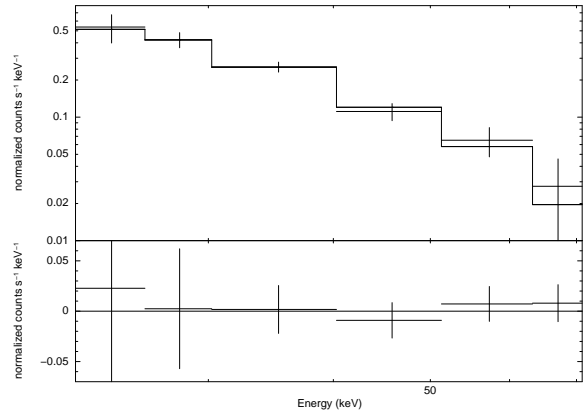}
\caption{IBIS/ISGRI spectrum of   AX~J1949.8+2534 (during revolution 1629) fitted by a black body. The lower panel shows
the residuals from the fit.}
\label{fig6}
\end{center}
\end{figure}
\subsubsection{IBIS/ISGRI detection in revolutions 1628 and 1629}

Renewed  hard X-ray activity from AX J1949.8+2534 was detected again by IBIS/ISGRI towards the end of Dec 2015. In fact, the source  was detected in the energy band 22--60 keV in both  revolutions n. 1628 ($\sim$ 7.1$\sigma$, $\sim$ 18 ks on-source exposure, ) and n. 1629 ($\sim$ 7.3$\sigma$,  $\sim$ 19 ks on-source exposure), spanning the time range from  2015 Dec 31 15:40 (UTC) to 2016 Jan 05 10:37 (UTC). The source showed no sign of flux variation on revolution timescale since the  22--60 keV measured average fluxes are fully consistent with each other within their uncertainties (10.6$\pm$1.5 mCrab and 9.5$\pm$1.3 mCrab, respectively). Unfortunately,  the source was not  in the IBIS/ISGRI FoV again throughout revolutions after n. 1629 so we cannot constrain the duration of this latest transient hard X-ray activity,   we can only   infer a lower limit  of $\sim$ 4 days. 

We stacked the data for  revolutions n. 1628 and 1629 with the aim of increasing the statistics of the detection.  AX~J1949.8+2534 was detected in the  mosaic  with a significance of 9.8$\sigma$ (22--60 keV) for a total on-source exposure of $\sim$ 37 ks.  No detection was obtained in the higher energy band 60--100 keV. The  22--60 keV measured average flux is 9.30$\pm$0.95 mCrab  which is fully compatible within the errors with that measured during the hard X-ray activity detected two months earlier  during revolution n. 1605. The source was never  in the JEM--X FoV during these latest  observations. 

We performed an investigation at  ScW level of both revolutions n. 1628 and 1629 with the aim of searching for possible flaring activity 
on short timescales (i.e. $\sim$ 2,000 s). Interestingly, we found  that  the source  was occasionally bright enough to be significantly detected even at   ScW level during the course of the observations. Table 3 lists  the  single ScWs during which  significant detections (i.e. $\ge$ 5$\sigma$) were achieved.  In particular,  we focussed our attention on the  four consecutive ScWs from n. 20 to 23 in revolution n. 1629 (they span a continuous temporal range of $\sim$ 3.5 hours) in order to extract an IBIS/ISGRI light curve with a fine temporal bin  of 200 s. As it can be seen from Fig. 2,   AX~J1949.8+2534 mainly shows an enhanced and rather constant flux with no major sign of flares. Only towards  the end of the light curve is there  sign of possibly a couple of short flares on $\sim$ 200 s timescale,  the strongest one (having a significance of 4.7$\sigma$) reached  a peak-flux of 180$\pm$38 mCrab or  (2$\pm$0.4)$\times$10$^{-9}$ erg cm $^{-2}$ s$^{-1}$ (22--60 keV). To establish if the source statistically  varied  during the entire light curve, we fitted it  with a constant and  applied  the $\chi^{2}$ test.  It was found a chance probability of  0.4 that this results is due to chance, i.e. the source is variable at only 60\% confidence level which reject  the hypothesis of variability. 
We have also made a mosaic of such  four consecutive ScWs, this yielded to a source detection of $\sim$ 
11$\sigma$ (22--60 keV). Given the good statistics, we extracted  an IBIS/ISGRI spectrum which was best fit by a black body 
($\chi^{2}_{\nu}$=0.9, 4 d.o.f., see Fig. 3) with  kT=8.6$^{+1.4}_{-1.2}$ keV and average  18--60 keV (20--40 keV) flux of  $\sim$ 6$\times$10$^{-10}$ erg cm$^{-2}$ s$^{-1}$ ($\sim$ 4$\times$10$^{-10}$ erg cm $^{-2}$ s$^{-1}$).  We note that both a thermal bremsstrahlung  ($\chi^{2}_{\nu}$=0.6, 4 d.o.f.) and a power law ($\chi^{2}_{\nu}$=0.75, 4 d.o.f)  provided a good description of the spectrum as well, with best fit parameter  values equal to   kT=34$^{+23}_{-11}$ keV and  $\Gamma$=2.5$\pm$0.5,   respectively. All such best fit parameter  values are consistent, within their uncertainties, with those obtained  from  the detection during revolution n. 1605.

\begin{table}
\caption {List of single ScWs in revolutions n. 1628 and 1629 during which significant source detections (i.e. $\ge$ 5$\sigma$) where achivied in the energy band 22--60 keV.} 
\label{tab:main_outbursts} 
\begin{tabular}{cccc}
\hline
\hline   
Telescope orbit    &     ScW        &     Flux                      &    Significance                         \\
    (n.)                   &       (n.)       &    (mCrab)               &   ($\sigma$)             \\
\hline    
1628                    &       30        &   66.5$\pm$10.2     &  6.5                   \\  
1629                    &       20        &   29.4$\pm$5.2       &  5.6                   \\  
1629                    &       21        & 22.8$\pm$4.2          &  5.4                   \\ 
1629                    &       22        & 24.8$\pm$4.3 &  5.8                   \\ 
1629                    &       23        &   28.5$\pm$5.1     &  5.6                   \\ 
1629                    &       33        &   76.1$\pm$13.4    &  5.7                   \\ 
1629                    &       35        &   47.6$\pm$9.2      &  5.2                   \\ 
\hline
\hline  
\end{tabular}
\end{table}

\subsubsection{IBIS/ISGRI refined position and upper limit}

In order to get the most  refined IBIS/ISGRI position of the source, we made a mosaic summing up all the three revolutions 1605, 1628 and 1629.   AX J1949.8+2534 was detected at 11.2$\sigma$ level (22--60 keV) with a total on-source exposure of $\sim$ 50 ks. 
Fig. 4 shows the corresponding significance map.  The best position  is  RA=297$^\circ$.49 and Dec=25$^\circ$.57 with a 90$\%$ confidence error circle radius equal to 2$'$.4.  

 AX J1949.8+2534 is not listed in the  latest published IBIS/ISGRI catalog  (Bird et al. 2016) despite extensive {\itshape INTEGRAL}  coverage of 
its sky region  ($\sim$ 2 Ms up to revolution n. 1000 considered in Bird et al. 2016) and  this information can be used to infer an upper limit on its persistent hard X-ray emission. By additionally considering the source exposure from  our present dataset ($\sim$ 0.25 Ms), we can infer a 3$\sigma$ upper limit of $\sim$ 0.4 mCrab or 3.2$\times$10$^{-12}$ erg cm$^{-2}$ s$^{-1}$ (20--40 keV) for persistent emission. When assuming the source peak flux as measured by IBIS/ISGRI from the outburst reported in section  
2.2.2, we can infer a dynamic range of $\ge$ 625

\begin{figure}
\begin{center}
\includegraphics[height=8cm,angle=0]{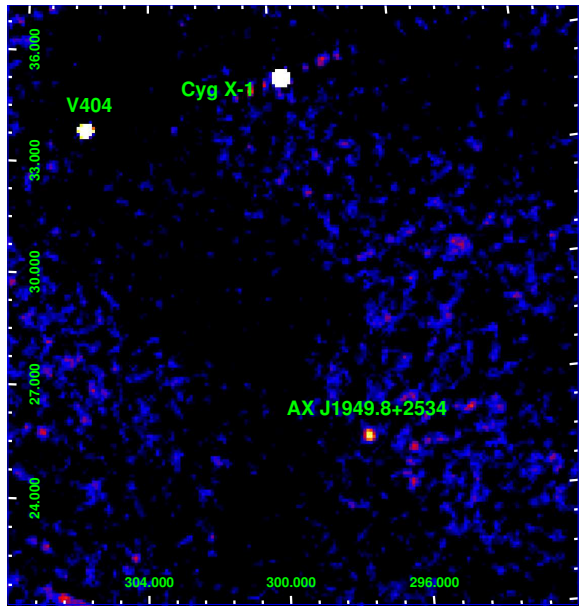}
\caption{IBIS/ISGRI significance mosaic map (22--60 keV) obtained by summing revolutions 1605, 1628, 1629. The source 
 AX~J1949.8+2534 is detected at $\sim$ 11$\sigma$ level. }
\label{fig6}
\end{center}
\end{figure}

\section{Soft X-ray observations}

\subsection{$Swift/XRT$}

Following the  two newly discovered  IBIS/ISGRI detections of  AX~J1949.8+2534 reported here, we triggered a ToO observation  of the sky region with the $Swift$ satellite  (Gehrels et al. 2004) with the main aims to i) refine the position of the source with a much higher accuracy; ii)  perform a spectral analysis in soft X-rays for the first time.  The observation was performed on  2016, April 25  (ID: 00034497001). Standard data reduction and analysis were perfomed using {\em HEASOFT} version 6.18 together with the most updated  {\em Swift}/XRT calibration  files.
The XRT data were reprocessed using {\sc xrtpipeline} (v0.13.2). 

The {\em Swift}/XRT (PC) observation resulted in a net exposure time of 2,929 s, and
it detected a faint X-ray counterpart ($\sim$5.3$\sigma$, 0.3--10 keV) within both the IBIS/ISGRI and {\itshape ASCA} error circles (see Fig. 5).
The best determined XRT position is at R.A. (J2000) = $19^{\rm h}49^{\rm m}55\fs19$ 
Dec (J2000) = $+25^\circ 33\arcmin57\farcs5$ with an error radius of 3$''$.7 (90$\%$ confidence) using the XRT--UVOT alignment and matching UVOT
field sources to the USNO-B1 catalog (see Evans et al. 2009 and  http:/www.swift.ac.uk/user$\textunderscore$objects).

We then extracted counts from a circle with a radius of 20 pixels centered on the source position, together with
a background from an annular region
centered on the source position, with inner and outer radii of 30 and 60 pixels, repectively.
The source net count rate in the energy range 0.3--10 keV was (8.08$\pm{1.71})\times10^{-3}$~counts~s$^{-1}$.

Given the  low statistics, we rebinned the spectrum to 1 count~bin$^{-1}$ and adopted Cash statistics (Cash 1979) in {\sc xspec}.
The {\itshape Swift/XRT} spectrum  was well fitted by an absorbed power law (C-Stat = 20.8,  24 d.o.f.)
where the absorption N$_{\rm H}$ was fixed at 1.18$\times$10$^{22}$~cm$^{-2}$ (the total Galactic absorption in the source direction, Willingale  et al. 2013). We obtained a photon index of $\Gamma$=0.2$\pm$0.9.
The 0.3--10 keV observed flux was 1.8$\times10^{-12}$~erg~cm$^{-2}$~s$^{-1}$ (2$\times10^{-12}$~erg~cm$^{-2}$~s$^{-1}$ corrected for the absorption).

\begin{table*}
\caption {List of NIR sources (as taken from the UKIDSS Galactic Plane Survey) located inside the 90\%  confidence  error circle (n. 1)  and 95\%  confidence  error circle (from n. 2 to n. 5)  of AX 1949.8+2534. $\dagger$  source also reported in the 2MASS  catalog as  J19495543+2533599. 
The table lists their  JHK magnitudes (lower limits  are derived according to Lawrence et al. 2007), offset from the XRT coordinates, Q value (see section 4), 
inferred spectral type,  reddening and distance}
\begin{tabular}{cccccccccc}
\hline
\hline   
n.    & name                               &    J          &   H          &       K         & offset      & Q  & Spectral Type & A$_v$ & d  \\
      &                                         &    (mag)          &   (mag)  &       (mag)    &                &    &                         & (mag)     & (kpc)  \\
\hline    
1     &  J194955.02+253354.6	  &     $>$19.9                   &     $>$19.0     &     17.788$\pm$0.142             &   3.66$''$   &    &   &  &    \\ 
\hline 
2     &  J194954.99+253400.2     &    17.094$\pm$0.015   &   16.182$\pm$0.013   &   15.743$\pm$0.022  & 3.78$''$  &  0.17  &   late type &  &   \\
3     & 	J194955.12+253401.4	  &    14.889$\pm$0.003   & 14.263$\pm$0.003     &   13.902$\pm$0.005     & 4.02$''$ & 0.01   & B0V  & 7.1 & 17.3 \\
4     &  J194955.31+253353.8  &     17.102$\pm$0.015        &   15.878$\pm$0.010           &      15.285$\pm$0.015        &   4.02$''$   &  0.22  &  K0V & 4.4 & 1.3  \\ 
5$\dagger$    &  J194955.42+253359.9	&       9.900$\pm$0.022  &           9.071$\pm$0.016   &      8.637$\pm$0.018        &   4.02$''$   & 0.09   & B0.5Ia  & 7.2 & 8.8 \\ 
\hline
\hline  
\end{tabular}
\end{table*} 

\subsection{$Chandra$}

The  ACIS High Resolution Camera   onboard  {\itshape Chandra} (Weisskopf et al. 2000) observed AX J1949.8+2534
on 2008 Feb 08  for a total exposure time of $\sim$ 1.16 ks. This targeted observation was performed in the context  of  
the ChIcAGO survey (Anderson et al. 2014),  aimed at classifying a selected list of unidentified X-ray sources discovered 
during the {\itshape ASCA} Galactic plane survey. No significant X-ray source was detected inside the entire  field of view pertaining to this {\itshape Chandra} observation,  the  ChIcAGO survey team reports only  a $\sim$2$\sigma$ excess (labelled as ChI 194951+2534$\_$1 from Table 1 in Anderson et al. 2014) located at 1$'$.6 from the {\itshape ASCA} position of AX J1949.8+2534. 
 We used the  count rate of ChI 194951+2534$\_$1 in order to estimate with WEBPIMMS  a 0.3--10 keV  observed (unabsorbed)  3$\sigma$  upper limit of  9.6$\times10^{-13}$~erg~cm$^{-2}$~s$^{-1}$ 
(1.05$\times10^{-12}$~erg~cm$^{-2}$~s$^{-1}$)  for  AX~J1949.8+2534. We assumed the same spectral model as from   the {\itshape Swift/ XRT} observation. This is a factor of about 2 lower than the X-ray flux measured during the  {\itshape Swift/XRT} detection.

\begin{figure}
\begin{center}
\includegraphics[height=6.85cm,angle=0]{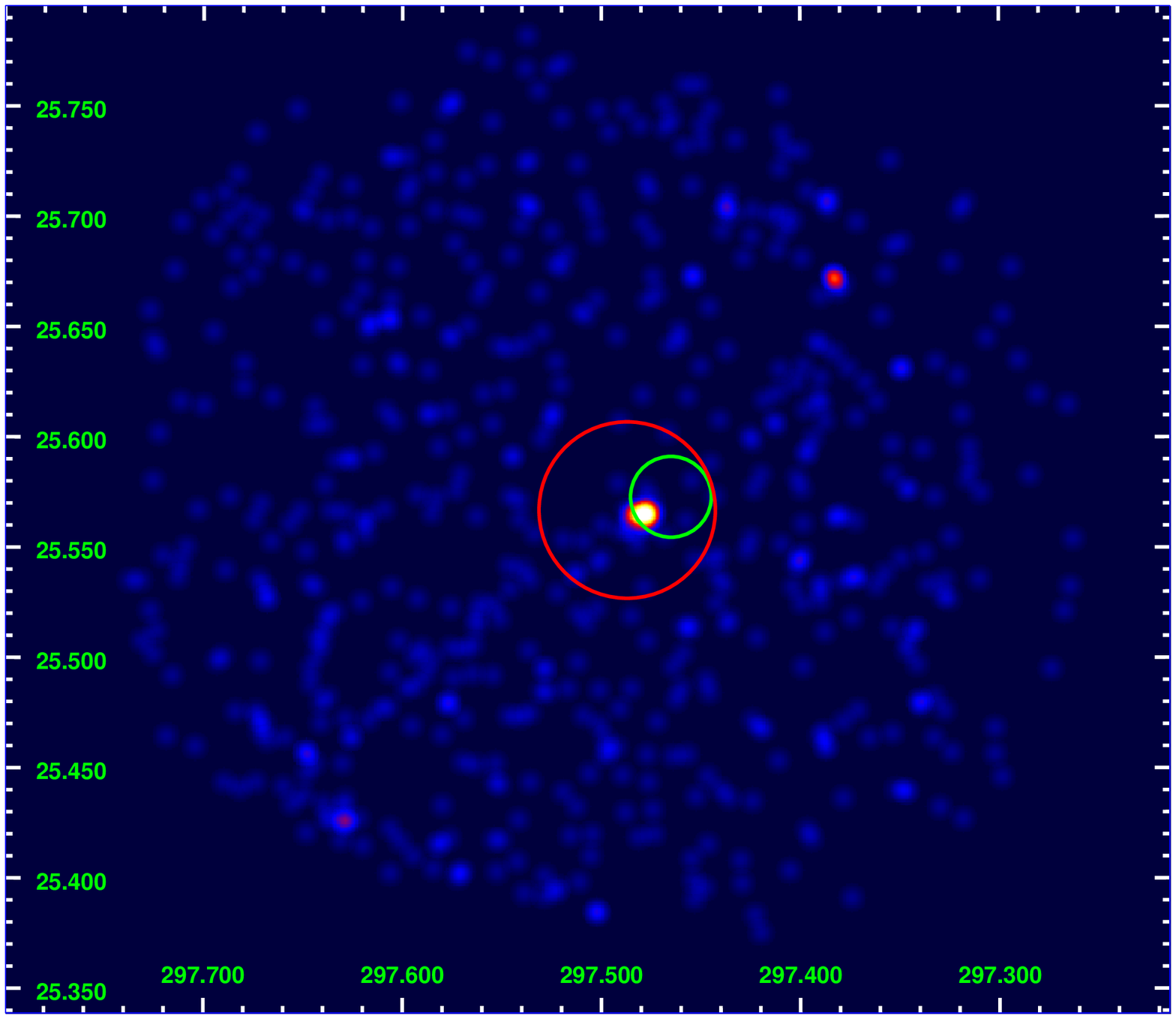}
\caption {{\itshape Swift/XRT} image (0.3--10 keV) with superimposed the {\itshape ASCA} (green) and IBIS/ISGRI (red) error circles having a radius of 1$'$ and 2$'$.4, respectively. AX~J1949.8+2534 is detected inside both error circles at $\sim$ 5.3$\sigma$ level.}
\label{fig6}
\end{center}
\begin{center}
\includegraphics[height=6.85cm,angle=0]{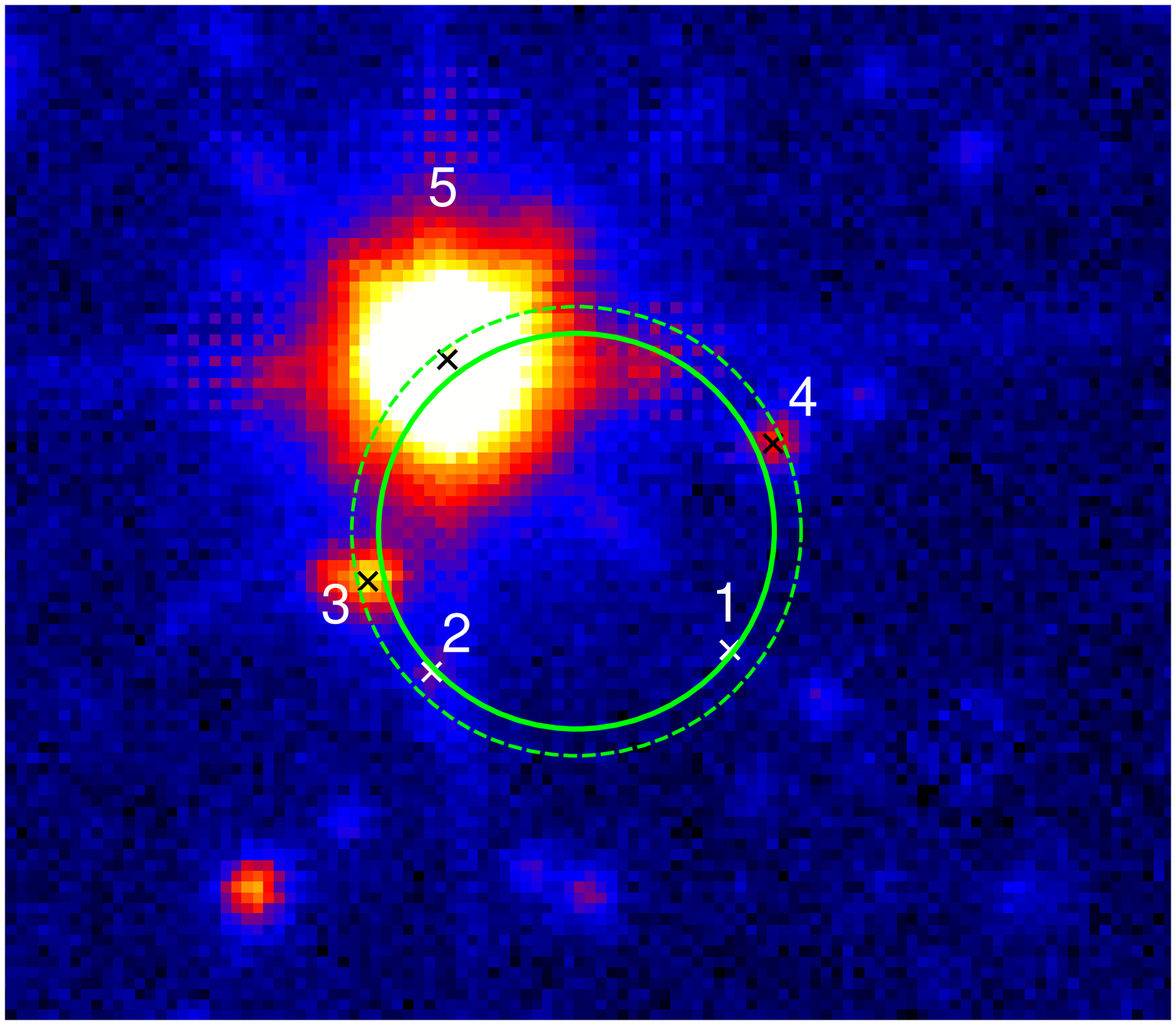}
\caption{UKIDSS image in the $K$ band (as downloaded from the UKIDSS archive, http://wsa.roe.ac.uk) with superimposed the {\itshape Swift/XRT}  error circles at 90$\%$ confidence 
(solid green,  radius of 3$''$.7)  and  95$\%$ confidence (dashed green,  radius of 4$''$.2)}
\label{fig6}
\end{center}
\end{figure}

\section{Search for infrared, optical and radio counterparts}

The identification of lower energy counterparts represents a mandatory step in determining the nature of unidentified Galactic X-ray sources. To this aim, we obtained with {\itshape Swift/XRT} a 90$\%$ confidence source positional accuracy of 3$''$.7  which significantly improves the  previously available
{\itshape ASCA} uncertainty of $\sim$ 1$'$.   This  allowed   us  to perform for the first time a search for 
counterparts from radio to optical bands,  by using all the available catalogs in the HEASARC database.  

No catalogued radio and optical source is located within the arcsecond sized {\itshape XRT} error circle. In the optical V band, a  lower  limit of V$>$21 can be inferred  from the USNO survey limit as reported in Monet et al. (2003).   Conversely, only one  near  infrared (NIR) source has been   detected by the UKIDSS Galactic Plane Survey (Lucas et al. 2008) inside the 90$\%$ confidence {\itshape XRT} error circle, such object  is  listed  in Table 4 as n. 1.  Fig. 6 shows 
both the {\itshape XRT} error circles   at 90$\%$ and 95$\%$ confidence, superimposed on the $K$ band UKIDSS image. We note that source n. 1  is located  at 3$''$.66 distance from the XRT coordinates centroid, i.e.  on the  90$\%$ confidence error circle (radius of  3$''$.7, solid circle). It is very  faint in the  $K$  band (magnitude of $\sim$ 17.8,  as such it is  not evident in the image) and it  is undetected in the $J$ and $H$ bands.  As a consequence  we  consider very unlikely the possibility that this extremely faint object  could be a reliable NIR counterpart of AX~J1949.8+2534. 
For the sake of completeness, in  Fig. 6  we note that  a few brighter infrared sources are  located    at $\sim$ 4$''$ distance from the {\itshape XRT}  centroid, i.e.  slightly  outside the  90$\%$ confidence error circle (radius of  3$''$.7, solid circle) and  well inside the 95$\%$ confidence {\itshape XRT} error circle radius of 4$''$.2 (dashed circle). Such objects  are listed in Table 4 with numbers from 2 to 5.

If we use the reddening-free NIR diagnostic $Q$ of Negueruela \& Schurch (2007) 
for the  objects n. 2 and n. 4  in Table 4, then 
we find that none of them  has a $Q$ value typical of early-type stars 
(i.e. $\leq$ 0). As a matter of fact, they show $Q$ values of 
0.17 and  0.22 respectively, which are much more similar to those of 
intermediate or late type stars. For example, if we consider  the brightest infrared object among the two (n. 4), then we find that it  is compatible with being a  star of spectral type   K0V  with a reddening of A$_v$=4.4 mag, located at  a distance of $\sim$ 1.3 kpc.  Its implied $V$ magnitude is $\sim$ 23.4, which is consistent with not detecting it in the V band  according to the USNO catalog.
As for the brightest infrared object  (n. 5), it is  also reported in the 2MASS  infrared source catalog (namely J19495543+2533599) with  magnitudes equal to $J$=9.900$\pm$0.022, $H$=9.071$\pm$0.016 and $K$=8.637$\pm$0.018 as well as in the  optical USNO--B1.0 catalog (1155--0421415) with magnitudes of $I$=12.87, $R2$=14.62, $B2$=17.35.  Notably, its  $Q$ value (0.09) is typical of early-type stars (Negueruela \& Schurch 2007). Moreover,   by  comparing its $K$ magnitude with the right panel of Fig. 1 in Reig \& Milonaki (2016) we note that  it is located inside the box populated  by blue supergiants.  In fact, we found  that the observed $K$ and $J$ magnitudes and the observed $J-K$ color are compatible with a B0.5Ia spectral type supergiant star, for a distance of $\sim$ 8.8 kpc  and a reddening of  A$_v$=7.2 mag. Its implied  apparent $V$ magnitude is $\sim$ 15, which is consistent with the  detection of the source as reported  in the  USNO catalog. Finally, also the infrared object n. 3 is characterized by a Q value (0.01)  typical of an early-type star. In this case, its observed NIR magnitudes are not compatible with being a  supergiant star  since this  would require an extremely large distance ($\sim$ 70 kpc, i.e. object outside the Galaxy) and an high extinction 
(A$_v$=6.7 mag). On the other hand, it is more compatible with  being a  main sequence B0V  spectral type  star, for a distance of $\sim$ 17.3 kpc  and a reddening of  A$_v$=7.1 mag,  the implied  apparent $V$ magnitude is $\sim$ 19.3, which is consistent with the  detection of the source as reported  in the  USNO catalog with a B magnitude equal to $\sim$ 19.

\section{Discussion}
We have presented mainly IBIS/ISGRI results on newly discovered hard X-ray 
activity from the unidentified transient AX~J1949.8+2534, the first ever emission to be reported above 20 keV. Hard X-ray outbursts have been detected twice, on November 2015  and January 2016,  respectively.   
Furthermore,  we point out that we have searched the entire currently available IBIS/ISGRI public data archive (revolution 25--1619, Paizis et al. 2013, 2017)  for possible additional  outbursts of AX J1949.8+2534. No detections have been found  at ScW level above a significance value of 7$\sigma$ in both energy bands 22--50 and 50--100 keV.  The source exposure time  obtained from the entire archive  is  of the order of $\sim$ 7   Ms.  The inferred duty cycle is as low as $\sim$  4$\%$. 

We can use all the collected multiwavelength data to consider the possible nature of AX~J1949.8+2534.

\subsection{ Low Mass X--ray Binary or  Cataclysmic Variable? }
As we noted before, two NIR objects,  compatible with being late type spectral stars,
are present  within the 95\% confidence {\itshape XRT} error circle (n. 2 and n. 4 in Table 4 and Fig. 6).  In principle, this could suggest  a  Low Mass X--ray Binary nature (LMXB) or alternatively a  Cataclysmic Variable one (CV). However, in the following we show that both such scenarios   suffer serious drawbacks when broad-band X-ray results are taken into account. 

The LMXB  hypothesis is incompatible with both the very hard  X-ray spectrum measured by {\itshape Swift/XRT} (i.e. $\Gamma$$\sim$0.2) as well as with the particularly short duration of the hard X-ray activity detected by IBIS/ISGRI ($\sim$ 1.5 days, in the only case when it was possible to firmly constrain the duration). 
In fact, transient LXMBs are know to display X-ray outbursts whose duration is typically of the order of weeks/months, they  are characterized by soft X-ray spectra below 10 keV (due to disk black body emission). From Table 4, we note that  the two  NIR objects  compatible with a late type nature (i. e. 
n. 2 and  n. 4) are particularly  weak. If we consider as possible counterpart the brightest among the two  (n. 4) then we showed that it is compatible with a late  type star located at $\sim$ 1.3 kpc. The corresponding average X-ray luminosities during
the two hard X-ray outbursts detected by IBIS/ISGRI are  of the order of $\sim$  2$\times10^{34}$~erg~s$^{-1}$.  This is way too low if 
compared to typical  luminous X-ray outbursts  from LMXBs (up to $\sim$  10$^{37-38}$~erg~s$^{-1}$). 
For the sake of completeness, we note that recently a growing number of LMXBs have been found to show outbursts reaching peak X-ray luminosity of only $\sim$ 10$^{34-36}$~erg~s$^{-1}$ in the soft X-ray band 2--10 keV. They belong to a more general class of X-ray transients dubbed as  Very Faint X-ray Transients (VFXTs), which  are believed to be the the faintest known X-ray accretors (Degenaar \& Wijnands 2009, 2011). If we extrapolate the IBIS/ISGRI spectral shape of AX~J1949.8+2534  in outbursts to the 2--10 keV X-ray flux, then 
we obtain a 2--10 keV X-ray luminosity (at 1.3 kpc) of the order of $\sim$ 2$\times10^{33}$~erg~s$^{-1}$, i.e. lower than that typical of VFXTs.  With all the above  information at hand a LMXB  nature for  AX~J1949.8+2534  seems to be not viable.

Alternatively, we are left with the CV  hypothesis. In principle, both the {\itshape Swift/XRT} and IBIS/ISGRI spectral X-ray characteristics are compatible with this scenario (Barlow et al. 2006, Landi et al. 2009). However, we note that to date all the CVs  detected by both
 IBIS/ISGRI and {\itshape Swift/BAT} above 20 keV are weak persistent hard X-ray  sources with typical luminosities  in the range $\sim$ 10$^{32-34}$~erg~s$^{-1}$ (Barlow et al. 2006, Revnivtsev et al. 2008, Brunschweiger et al. 2009).  No CV has never been detected as a transient hard X-ray source. This is completely at odds  with the short transient behavior  of AX~J1949.8+2534 as observed by IBIS/ISGRI.  
Moreover, if we consider the brightest  NIR object  compatible with a late type nature (n. 4), then it is compatible with a late type M5V main sequence star (which is typical of CVs, Smith et al. 1998) located at $\sim$ 230 pc. Consequently, the IBIS/ISGRI upper limit on the persistent hard X-ray emission of AX~J1949.8+2534  would translate into a 20--40 keV luminosity of $\leq$ 2$\times10^{31}$~erg~s$^{-1}$, i.e. significantly lower than typical measurements obtained to date with both IBIS/ISGRI and {\itshape Swift/BAT}. The same holds for the {\itshape Swift/XRT} detection whose soft X-ray flux would translates into a luminosity of $\sim$ 1.2$\times10^{31}$~erg~s$^{-1}$. All the above results point to largely disfavor the CV interpretation for  AX~J1949.8+2534.

\subsection{High Mass X-ray Binary}
The location of the source on the Galactic plane (b$\sim$--0.3$^{\circ}$) and  both {\itshape Swift/XRT} and IBIS/ISGRI spectral  characteristics (especially the hardness of the soft  X-ray spectrum, i.e. $\Gamma$$\sim$0.2) are fully compatible with a High Mass X-ray Binary nature (HMXB). In particular, the hard X-ray transient behavior (e.g. dynamic range $\ge$ 620  and duration of $\sim$1.5 days and $\ge$ 4 days from the two IBIS/ISGRI detected activities, respectively) could  be typical of the  Be HMXB class. In addition we note that mentioned  X-ray characteristics are compatible as well with a Supergiant Fast X-ray Transients nature (SFXTs, Sguera et al. 2008), which are a newly discovered  class of HMXBs (Sguera et al. 2005, 2006, Negueruela et al. 2005)

Although classical SFXTs usually display above 20 keV hard X-ray outbursts lasting much less than a day,  a few other SFXTs are known to show unusually longer hard X-ray activity,  exceptionally lasting several days (e.g.  IGR~J18483$-$0311, Sguera et al. 2015;  IGR J17354$-$3255, Sguera et al. 2011),  i.e.  comparable to the duration of the hard X-ray activity detected from AX~J1949.8+2534.   

 We must note that, in principle,  this proposed HMXB interpretation could suffer some drawbacks when
NIR data are  combined with the  90\% confidence {\itshape XRT} positional uncertainty. In fact, searching within the latter, we  pinpointed only one very faint NIR object (n. 1 in Table 4), whose magnitudes and colors are not compatible with being an early type spectral star, i.e. at odds with an HMXB nature. However,
slightly  outside the 90\% confidence {\itshape XRT} error circle (radius of  3$''$.7) we note the presence of two bright NIR sources (n. 3 and n. 5) located at 4$''$ distance from the XRT centroid, i.e. well inside the 95\% confidence {\itshape XRT} positional uncertainty (radius of 4$''$.2). Both have observed $K$ and $J$ magnitudes and  $J-K$ color compatible with an early type nature. 

Specifically, the characteristics  of the  NIR object n. 3 are compatible with  a main sequence  early type spectral star (B0V), supporting  a Be HMXB scenario. If we consider its  calculated  distance of $\sim$ 17.3 kpc, then the two hard X-ray outbursts  detected by IBIS/ISGRI  from AX J1949.8+2534 would have a 18--60 keV  average (peak) luminosity of 4$\times10^{36}$~erg~s$^{-1}$ (2$\times10^{37}$~erg~s$^{-1}$), both such values are typical of periodic type I  X-ray outbursts from Be HMXBs.  As for the soft X-ray band, the measured {\itshape Swift/XRT} flux translates into a luminosity of $\sim$ 6$\times10^{34}$~erg~s$^{-1}$.  Such value is not representative of the  X-ray quiescence of Be HMXBs which is usually much lower. Conversely,  it could be more typical of quasi-persistent Be HMXBs which are  a  small subclass characterized by a persistent  low X-ray luminosity of $\sim$ 10$^{34}$~erg~s$^{-1}$, varying by up to a factor of $\sim$ 10 at most, hence not displaying type I or II outburst behavior (see Reig $\&$ Roche 1999).  We note that this latter characteristic is at odds with the bright  outburst behavior detected by IBIS/ISGRI from AX~J1949.8+2534, as such it represents a significant drawback for the Be HMXB scenario.

As for the association with the brightest NIR object n. 5, their characteristics  are fully compatible with being an early type spectral star of supergiant nature (B0.5Ia). This would fully support  an SFXT   scenario.  We  are  aware that, in principle, to extend the search for NIR counterparts outside the canonical  90\% {\itshape XRT} positional uncertainty could  eventually be a  dangerous  approach,  because  of  the  possibility that unrelated NIR sources could be included  and mistakenly assumed as counterparts. Bearing this in mind, we  took  into  account  the  possibility  that  such  an  association could be simply a chance coincidence and accordingly calculated the  probability of a random association with an infrared source having  magnitude $K$ $\leq$ 8.6 as given by $P$ = 1 - e$^{-\pi\rho d^2}$   (where d is the distance between the {\itshape XRT} centroid and the associated NIR  candidate, and $\rho$  is the local spatial density of NIR sources computed in an area of a few degree radius around the candidate).  We estimated a probability of 0.1\%, i.e.  0.001 chance coincidence  is expected. Such  particularly low probability   strongly supports a real  physical association, hence the viable possibility that  this  infrared object is a reliable  counterpart of the unidentified X-ray source AX~J1949.8+2534.  If we consider the calculated  distance of $\sim$ 8.8 kpc in the case of a B0.5Ia spectral type  supergiant for the proposed NIR counterpart, then the two hard X-ray outbursts  detected by IBIS/ISGRI  from AX J1949.8+2534 would have a 18--60 keV  average (peak) luminosity of 1.1$\times10^{36}$~erg~s$^{-1}$ (5.5$\times10^{36}$~erg~s$^{-1}$).   From archival {\itshape INTEGRAL/IBIS} observations, we placed a 3$\sigma$ upper limit of  $\sim$ 3$\times10^{34}$~erg~s$^{-1}$ for the persistent hard X-ray luminosity.  Both such values are very similar to those of known confirmed SFXTs. As for the soft X-ray band, the measured {\itshape Swift/XRT} flux translates into a luminosity of $\sim$ 2$\times10^{34}$~erg~s$^{-1}$, which  is fully compatible with the so-called  intermediate intensity X-ray state during which typical SFXTs spend the majority of
their time (Sidoli et al. 2008).  In this context, we note that the  prolongated  duration of the hard X-ray activity discovered from  AX~J1949.8+2534 ($\sim$ 1.5 and $\ge$ 4 days, respectively) is at odds with the much shorter durations typically marking outbursts from classical SFXTs  above 20 keV (i.e. a few hours). A similar characteristic was previously reported only for a few other SFXT sources. Our new findings on AX~J1949.8+2534  could strengthen the  idea that unusually long hard X-ray outbursts would not be particularly exceptional among the class of SFXTs.
           
\section{Conclusions}
We reported on the IBIS/ISGRI discovery of two new hard X-ray outbursts from the unidentified transient AX~J1949.8+2534, the first ever emission to be detected  above 20 keV.   A follow-up  observation  of the sky region with the $Swift$ satellite allowed for the first time to perform  a  soft X-ray spectral analysis  as well as significantly improve the positional uncertainty to arcsecond size. This permitted us to pinpoint two bright infrared sources as most likely  candidate counterparts. Both are compatible with being  early type spectral stars hence supporting  an HMXB nature, specifically an SFXT (more viable) or alternatively a Be HMXB.  Further detailed NIR or optical spectroscopy is mandatory to confirm their putative  supergiant and Be nature, respectively.  
Unfortunately, we can go no further on this issue because the current X-ray positional  uncertainty (3$''$.7) prevents us from unambiguously pinpointing
the correct single NIR counterpart.  Additional X-ray observations of AX~J1949.8+2534 using {\itshape Chandra}, for example, are strongly needed in order to achieve a finer position with an associated  smaller error circle.

\section*{Acknowledgments}
We thank the anonymous referee for useful comments which helped us to improve the quality of this paper.
We thank the {\em Swift} team, the PI, the duty scientists and science planners
 for making the two ToO observations reported here possible.  
We acknowledge financial support from the Italian Space Agency via INTEGRAL
ASI/INAF agreement n. 2013-025.R.1, and the grant from PRIN-INAF 2014, 
``Towards a unified picture of accretion in High Mass X-Ray Binaries''.
This work has made use of the INTEGRAL archive developed at INAF-IASF Milano (http://www.iasf-milano.inaf.it/\textasciitilde{}ada/GOLIA.html). 
This  research has made use of data and/or software provided by the 
High Energy Astrophysics Science Archive Research Center (HEASARC), which is a 
service of the Astrophysics Science Division at NASA/GSFC and the 
High Energy Astrophysics Division of the Smithsonian Astrophysical Observatory.

{}

\end{document}